\documentclass[twocolumn,aps,prb,floatfix]{revtex4-1}

\usepackage{graphicx}
\usepackage{epstopdf}
\usepackage{amsmath}
\usepackage{latexsym}
\usepackage{amsmath}
\usepackage{esdiff}
\usepackage{xcolor}
\usepackage{mathtools}
\usepackage{extarrows}

\newcommand{\beq}{\begin{equation}}
\newcommand{\eeq}{\end{equation}}

\newcommand{\eqWBL}{\xlongequal{\text{WBL}}}

\begin{document}

\title{Nanoscale functionalized superconducting transport channels as photon detectors}

\author{Catalin D. Spataru}
\thanks{Corresponding author.\\ cdspata@sandia.gov}
\author{Fran\c{c}ois L\'{e}onard}

\affiliation{Sandia National Laboratories, Livermore, California 94551, USA}

\begin{abstract}
Single-photon detectors have typically consisted of macroscopic materials where both the photon absorption and transduction to an electrical signal happen. Newly proposed designs suggest that large arrays of nanoscale detectors could provide improved performance in addition to decoupling the absorption and transduction processes. Here we study the properties of such a detector consisting of a nanoscale superconducting (SC) transport channel functionalized by a photon absorber. We explore two detection mechanisms based on photo-induced electrostatic gating and magnetic effects. To this end we model the narrow channel as a one-dimensional atomic chain and use a self-consistent Keldysh-Nambu Green's function formalism to describe non-equilibrium effects and SC phenomena. We consider cases where the photon creates electrostatic and magnetic changes in the absorber, as well as devices with strong and weak coupling to the metal leads.
Our results indicate that the most promising case is when the  SC channel is weakly coupled to the leads and in the presence of a background magnetic field, where photo-excitation of a magnetic molecule can trigger a SC-to-normal transition in the channel that leads to a change in the device current several times larger than in the case of a normal-phase channel device.
\end{abstract}

\maketitle

\section{Introduction}
The most efficient single-photon detectors are currently based on superconducting (SC) nanowires \cite{Goltsman, Natarajan,Karl}. A single-photon absorption event in the superconductor is followed by a sequence of relaxation processes involving electron-electron and electron-phonon interactions, which culminate in the creation of a hot spot that triggers a phase transition to the normal state \cite{Allmarass}. Bypassing the energy cascade associated with such photo-induced temperature effects is desirable because it reduces the time scale for detection and the associated jitter. Besides temperature, there are other effects that directly impact the SC state such as electro-strictive, electrostatic gating and magnetic interactions. The question we seek to answer is whether such effects can be exploited for single-photon detection.

In parallel, there has been interest in developing new detector architectures that can overcome the limitations of existing systems. Along those lines, it was recently proposed that dense arrays of nanoscale detector elements could provide significant performance improvements \cite{Young}. One embodiment of such detector elements consists of functionalization nanoscale electronic transport channels functionalized with molecules, quantum dots, or other photon absorbers \cite{Leonard_SciRep, Bergemann}. We recently studied the detailed quantum dynamics of single-photon detection in a functionalized {\it semiconducting} electronic transport channel \cite{Spataru_singlePh} and found that a large signal-to-noise-ratio can be achieved for ultra-short single-photon pulses. An open questions is whether a SC electronic transport channel could be more efficient at detecting single photons, allowing to also bypass the thermal energy cascade.

Here we combine the Keldysh non-quilibrium Green's function formalism \cite{Keldysh} with the Nambu description \cite{Nambu} of the superconductor to study the change in electronic transport through a SC channel upon absorption of a photon by a nearby absorber. We consider the cases where the absorber undergoes changes in its permanent electric dipole or magnetic moment. In both cases the direct electromagnetic interaction between the absorber and the SC channel bypasses the usual thermal energy cascade. We show that under certain conditions photo-induced magnetic perturbations can trigger a SC-to-normal transition in the channel which in turn leads to a detectable event.

\section{Device system} 

\begin{figure}
\includegraphics[trim=-20 0 -20 0,clip,width=\columnwidth]{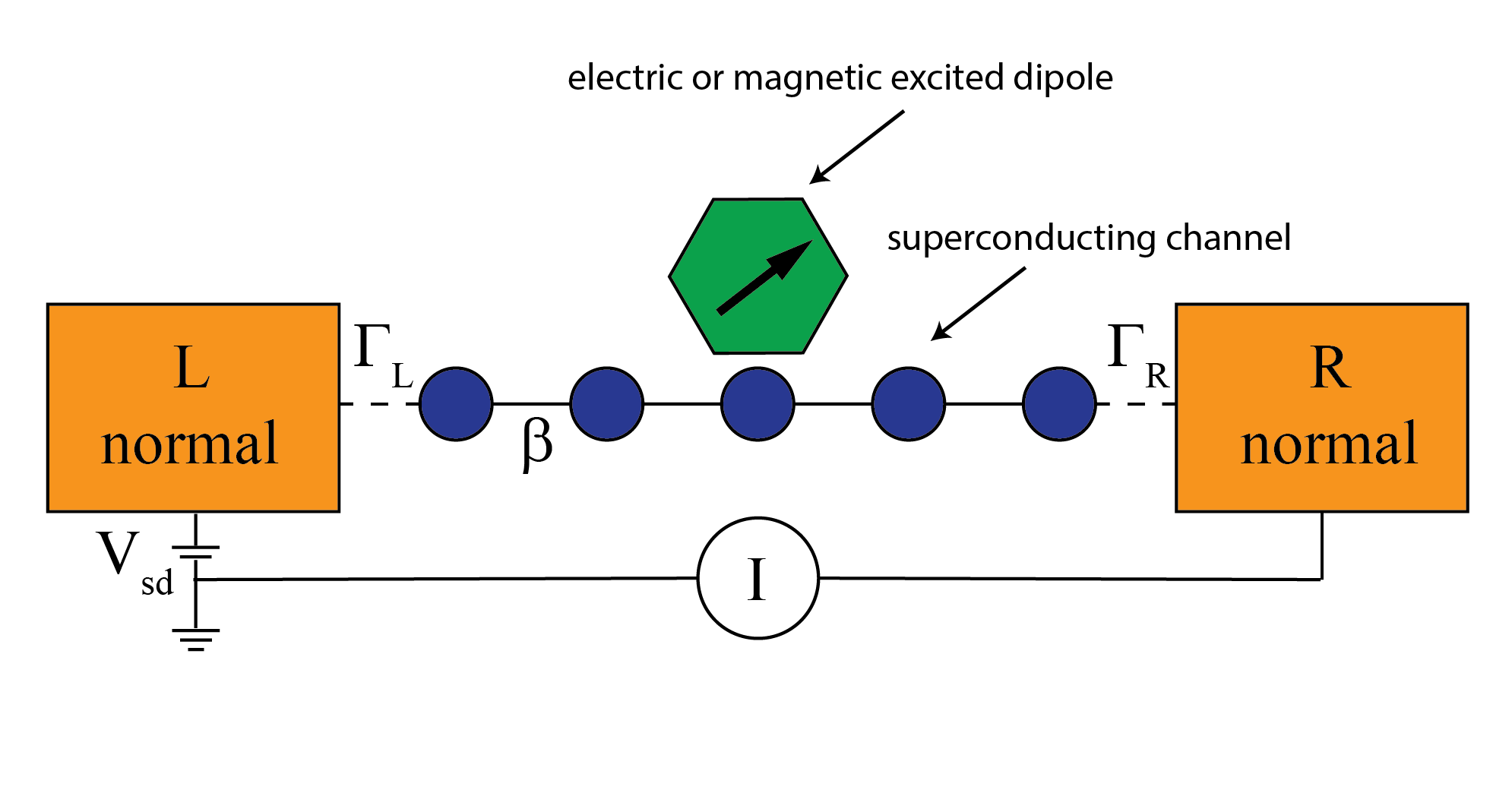}
\caption{Sketch of a device consisting of a superconducting quantum transport channel functionalized with a photon absorber. The superconductor is connected to normal-metal leads, between which a current is measured. The relevant tight binding parameters are indicated with greek letters.}
\vspace{-0.5cm}
\label{sketch}
\end{figure}

We consider a normal-superconductor-normal (NSN) device configuration with the device geometry sketched in Fig.~\ref{sketch}. It consists of a SC one-dimensional channel contacted by two normal metal electrodes. In practice the channel could be a nanowire or a nanotube \cite{CNT_SC,WS2_NT} which have been shown to have SC transition temperatures in the 5-15K range, but to reduce computational cost while capturing essential physics, here we model it as an atomic chain within the tight-binding approximation. We focus on a single band since the second subband is typically much higher in energy than the SC gaps considered here.
As opposed to typical SC nanowire single-photon detectors, photon absorption does not take place in the channel but by an absorber placed in close proximity to the SC channel. 

We assume that the photon energy is resonant with the absorber while the channel is not sensitive to the incoming light pulse (e.g., a small-diameter CNT has well-separated optical absorption peaks due to excitonic effects; we assume that the absorber optical gap is not matched to these peaks. We also assume that the system operates in the strong focusing regime where the light is concentrated on the absorbing element. This increases the light-absorber interaction and also helps to reduce the effects from absorption by the contacts.
The absorber (represented in Fig.~\ref{sketch} by the hexagon) could be a single-molecule or a larger quantum-dot that upon photo-excitation acquires a permanent electric or magnetic dipole, possibly via a photo-induced phase transition \cite{NewRef}. In the present work we do not calculate the absorber photoexcitation probability; rather we assume that the absorber has been photoexcited and we calculate the impact of the excited dipole on the electronic transport properties of the channel. 

As possible detection mechanisms we consider either i) photo-induced  electrostatic gating effects -e.g. due to a change of the absorber permanent electric dipole upon photo-excitation, or ii) photo-induced magnetic effects -e.g. due to a change in the absorber magnetic moment upon photo-excitation. 
Both types of perturbation can change the electronic density of states (DOS) at the Fermi level when the channel is in the normal phase and thus are expected to affect the SC state as well. In particular it has been discovered that monolayers of paramagnetic molecules can influence the temperature at which an adjacent thin layer of material becomes SC \cite{magM1,magM2,magM3}. In addition, photon-induced changes in molecular magnetic moments have been observed \cite{MM_ph}. Changes in molecular electric dipole moments are well-known to occur upon photon absorption, and have been shown to modulate semiconducting electronic transport channels \cite{Leonard_SciRep}. Their impact on SC channels may be expected based on the report of SC transistors where the gate is {essentially} equivalent to an electrostatic dipole \cite{SC_transistor}.

\section{Theoretical framework}
\subsection{Tight-binding Hamiltonian} 

The device is described within the grand-canonical picture using a Hamiltonian partitioned into left ($H^L$) and right ($H^R$) leads and channel ($H^C$) regions:
\beq
H=H^L-\mu_L N^L+H^R-\mu_R N^R+H^C\label{H_tot}
\eeq
where $N^{L,R}$ are the electron number operators in the left (L) and right (R) regions while the difference between left and right lead chemical potentials $\mu_L-\mu_R$ defines the applied bias.

The channel Hamiltonian is described using a tight-binding basis set with indices $\{m,n\}$:
\begin{multline}
H^C=-\beta \sum_{m,n,\sigma} c^\dagger_{m\sigma} c_{n\sigma} - U \sum_m n_{m\uparrow} n_{m\downarrow} +V_G \sum_{m,\sigma} n_{m\sigma} \\
+ \frac{\mu_B}{2} \sum_{m} (B_0+B^{abs}_m)(n_{m\uparrow}-n_{m\downarrow}).
\label{H_ch}
\end{multline}

The first term describes the bare channel with no electron interaction effects, with the nearest neighbor hopping integral set to $\beta = 0.5$ eV and an inter-site distance set to $a=5$ bohr (the summation over $n$ is restricted to $n=m\pm1$). The number of sites in the transport channel is denoted by $N$. $\beta$ 

The second, Hubbard-like local term includes electron correlation effects via an attractive contact interaction $-U<0$ corresponding to a s-wave type SC pairing potential. Here $n_{m\sigma}=c^\dagger_{m\sigma} c_{m\sigma} $ is the local number operator for an electron with spin $\sigma$ that can be either up ($\uparrow$) or down ($\downarrow$). 
We do not consider Coulomb interaction/electron correlation effects in the channel when the channel is in the normal state. The impact of repulsive Coulomb interaction effects could be important in certain device parameter regimes where exotic effects such as the Kondo effect may become important \cite{Hewson,Wang,SpataruKondo} . Including these effects is however beyond the scope of the present work. 
 
The third term describes an applied gate voltage $V_G$ acting uniformly along the channel. Magnetic effects are included via the last term which removes local spin-degeneracy via the Zeeman effect and accounts for: i) a uniform background magnetic field $B_0$ perpendicular to the chain, and ii) a non-uniform magnetic field $B^{abs}$ generated by an excited absorber with a permanent magnetic dipole $\mu_{abs}$ perpendicular to the chain. 

Finally, the coupling between leads and channel is assumed L-R symmetric and characterized by the induced broadenings $\Gamma_L=\Gamma_R\equiv\Gamma$. The SC pairing potential is zero inside the leads.  

\subsection{NEGF approach}

To describe non-equilibrium effects in conjunction with SC phenomena we employ the Keldysh-Nambu non-equilibrium Green's function (NEGF) formalism {(see the Appendix for details including the definition of the 4-component Nambu spinors.)}. NEGF-based studies of non-equilibrium phenomena in NSN junctions have employed both self-consistent \cite{Lambert,Yeyati1,Yeyati2} and non-self-consistent \cite{Claro,Tuovinen1,Tuovinen2} approaches. Here 'self-consistent' refers to the determination of the order parameter $F_{m}$. 
In this work $F_{m}$ is calculated rigorously via a self-consistent approach, directly from the off-diagonal Nambu component of the lesser Green's function:
$
F_{m}=i<c_{m \uparrow} c_{m \downarrow}> 
$.
The self-consistent process starts with a symmetry-broken initial guess {\it e.g.} by setting the order parameter $F_m$ to a constant. Convergence is facilitated by the use of the Pulay scheme \cite{Pulay} to mix the order parameter using previous iterations solutions \cite{Spataru_AM1}. Convergence is typically achieved in hundred iterations when $|F_m^{out}-F_m^{in}| < 10^{-13}$. 

We present briefly the NEGF expression for the expectation value of the current $I$ through the device (see the Appendix for {derivation}). The current can be decomposed into the normal  transmission{/single-electron tunneling} ($I_{N{T}}$), Andreev reflection ($I_{AR}$) and cross-Andreev ($I_{CA}$) components \cite{Andreev,Blonder,Devoret,Falci,CAR_L} of the current:
$
I=I_{N{T}}+I_{AR}+I_{CA}
$.
For the device parameters considered in this work the cross-Andreev component is negligible and it is sufficient to focus on the first two components, namely normal transmission \cite{Wingreen}
\begin{multline}
I_{N{T}} = 2 {\frac{e}{h}} \int d E [f(E-\mu_L)-f(E-\mu_R)] \\
\times Tr\{\Gamma^L G^r_{11}(E) \Gamma^R G^a_{11}(E)\}
\end{multline}
and Andreev reflection \cite{Andreev}
\begin{multline}
I_{AR} = 2 {\frac{e}{h}}  \int d E [f(E-\mu_L)-f(E+\mu_L)] \\
\times Tr\{\Gamma^L G^{r}_{12}(E) \Gamma^L G^{a}_{21}(E)\}
\label{I_AR}
\end{multline}
where $f$ is the Fermi-Dirac distribution function $f(E)=[exp(E/k_BT) + 1]^{-1}$ and $G^{r/a}_{ij}$ refers to the $ij$ component (in Nambu space) of the retarded/advanced channel Green's functions. The diagonal Nambu component is also used to obtain the electronic density of states projected on the channel: 
\beq
DOS(E)=-\frac{1}{\pi} Im\frac{Tr\{G^r_{11}(E)\}}{N}. 
\label{DOS}
\eeq

We note that the chemical potentials $\mu_{L,R}$ are referenced w.r.t.~the chemical potential of the SC condensate $\mu_C$. While in general in the presence of superconductivity ($U\ne 0$) $\mu_C$  needs to be determined self-consistently by imposing current conservation \cite{Lambert} along the channel, for the particular case where the total system has electron-hole symmetry $\mu_C$ is simply equal to the middle of the electronic band of either lead, {\it i.e.} it can be set to $0$.

{In our simulations we consider devices where the channel is SC ($U\ne0$) and undergoes a transition to a normal state as well as channels that are always in the normal state ($U=0$). We label the currents with the superscripts SC and N to denote these two cases. As above, subscripts are reserved for the components of the current when the channel is subject to superconductivity.}
 
\section{Results}

\subsection{Good metal contacts/strong channel-electrodes coupling}
\begin{figure}
\includegraphics[trim=-0 0 -0 0,clip,width=\columnwidth]{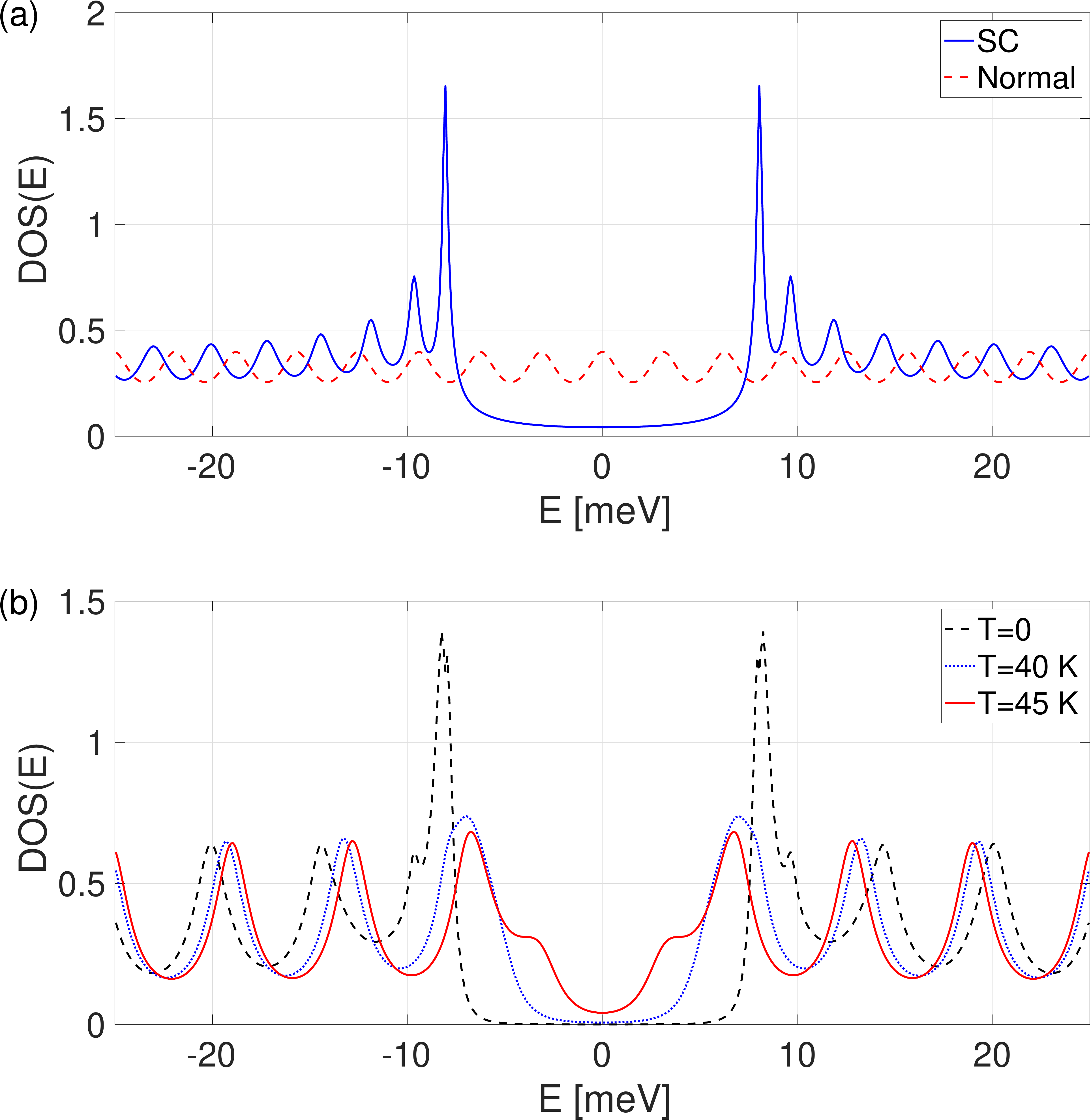}
\caption{Calculated DOS: a) at $T=0$, averaged over all channel sites{, for normal and SC case}. b) several temperatures, projected on the mid-channel site{, for the SC case}. Device parameters: $N=1001$, $\Gamma=0.25$ eV, $V_{sd}=0$, $U=-0.5$ eV. 
}
\vspace{-0.0cm}
\label{fig_DOS_N1001}
\end{figure}

First we consider the case of good metal contacts between source/drain electrodes and the transport channel. This is established by setting the coupling between leads and channel $\Gamma$ of the same order of magnitude as the hopping parameter inside the channel $\beta$, and as a representative value we choose $\Gamma = \beta/2$. 
We consider a channel with $N=1001$ and tune the gate voltage $V_G=0$ such that at zero source-drain bias voltage $V_{sd}=0$ the Fermi level pins the middle of the channel electronic band (which has a bandwidth of $\pm 1$ eV). In the normal (N) phase, the DOS is flat near the Fermi level, with small oscillations due to the fact that the energy spectrum is discrete. Upon turning on the attractive contact interaction term $U=-0.5$ eV, the channel becomes SC and a quasiparticle gap opens up in the DOS near the Fermi level, as seen in Fig.~\ref{fig_DOS_N1001}(a). The finite value of DOS near the Fermi level is due to the contribution from sites near the normal metal contacts. The channel is long enough that towards the middle of the channel the gap opening is nearly perfect. We note that at zero temperature $T=0$ and zero bias voltage $V_{sd}=0$ the SC gap is about $15$ meV; 
while this leads to a relatively large critical temperature -as discussed later-, it also helps in reducing the number of iterations needed to reach convergence - each iteration during the self-consistent loop is computationally demanding given the large $N$.

\begin{figure}
\includegraphics[trim=-0 0 -0 0,clip,width=\columnwidth]{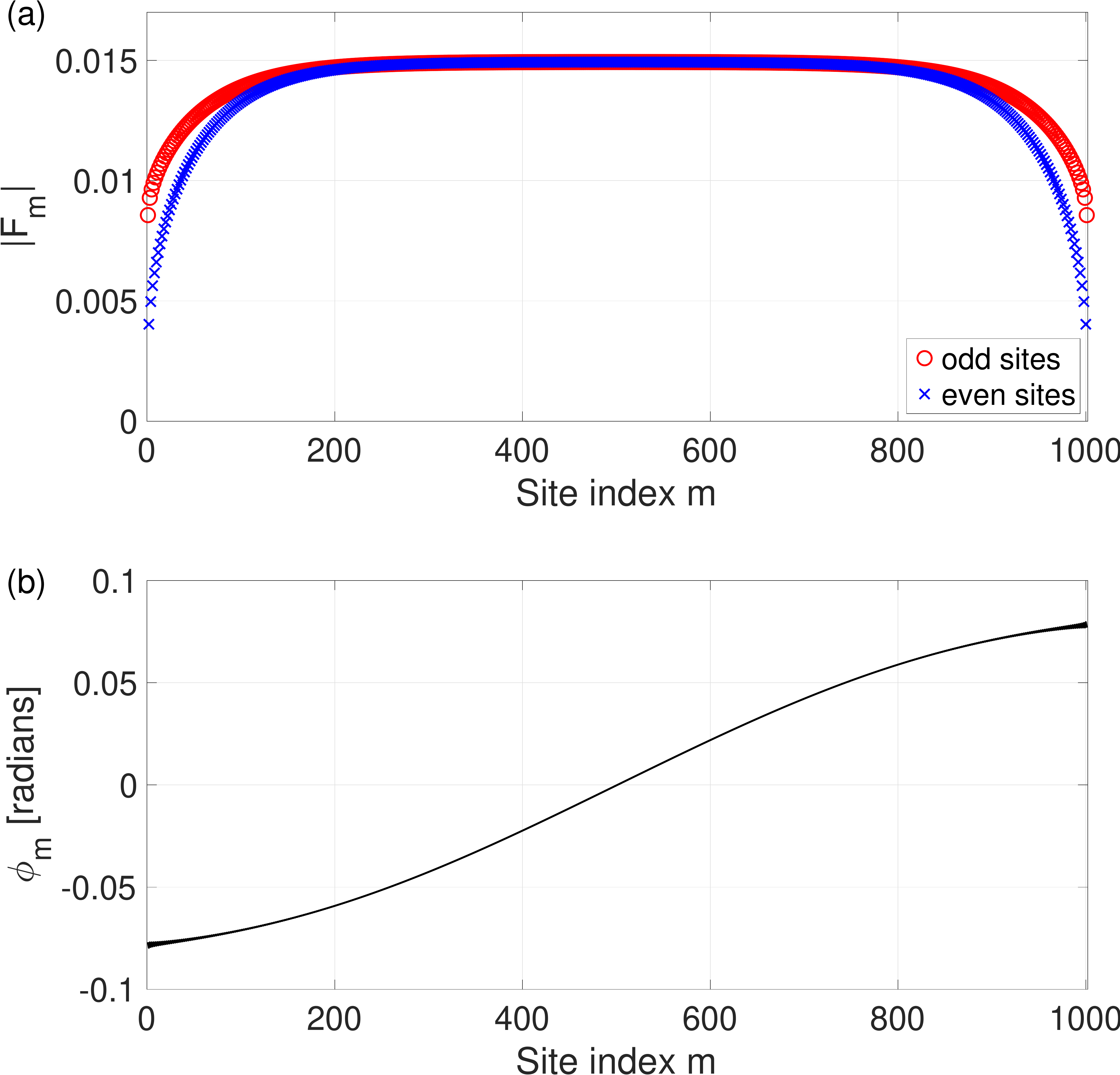}
\caption{Channel SC order parameter $F_m$. a) absolute value and b) phase. Device parameters: $N=1001$, $\Gamma=0.25$ eV, $T=0$ K, $V_{sd}=0.1$ meV, $U=-0.5$ eV. }
\label{fig_Fm_N1001}
\vspace{-0.0cm}
\end{figure}

Figure \ref{fig_Fm_N1001} shows the channel SC order parameter $F_m \equiv |F_m| e^{i\phi_m}$ at $T=0$ and small bias voltage $V_{sd}=0.1$ meV. $|F_m|$ is not impacted significantly by the small bias voltage, but is sensitive to the proximity of the normal-phase contacts and thus decreases in value near the ends of the channel. One notes in Fig.~\ref{fig_Fm_N1001}(a) that $|F_m|$ is asymmetric between even and odd sites. This can be traced back to the normal phase where electron wavefunctions associated with eigenchannels with energies near the middle of the channel electron band show strong asymmetry; in particular the eigen-channel with zero energy carries its entire weight on the odd sites. Figure \ref{fig_Fm_N1001}(b) shows the phase $\phi_m$ which displays a gradient  that increases with the bias voltage/current. Within more approximate theories where $|F_m|$ is assumed to be a constant independent of $m$, the phase gradient is directly related to the momentum $q$ of the Cooper pairs:  $q=\frac{1}{a}\frac{\partial\phi_m}{\partial m}$.  A non-zero $q$ simply corresponds to the flow of the SC condensate which carries a finite current when a bias voltage is applied.

\begin{figure}
\includegraphics[trim=-0 0 -0 0,clip,width=\columnwidth]{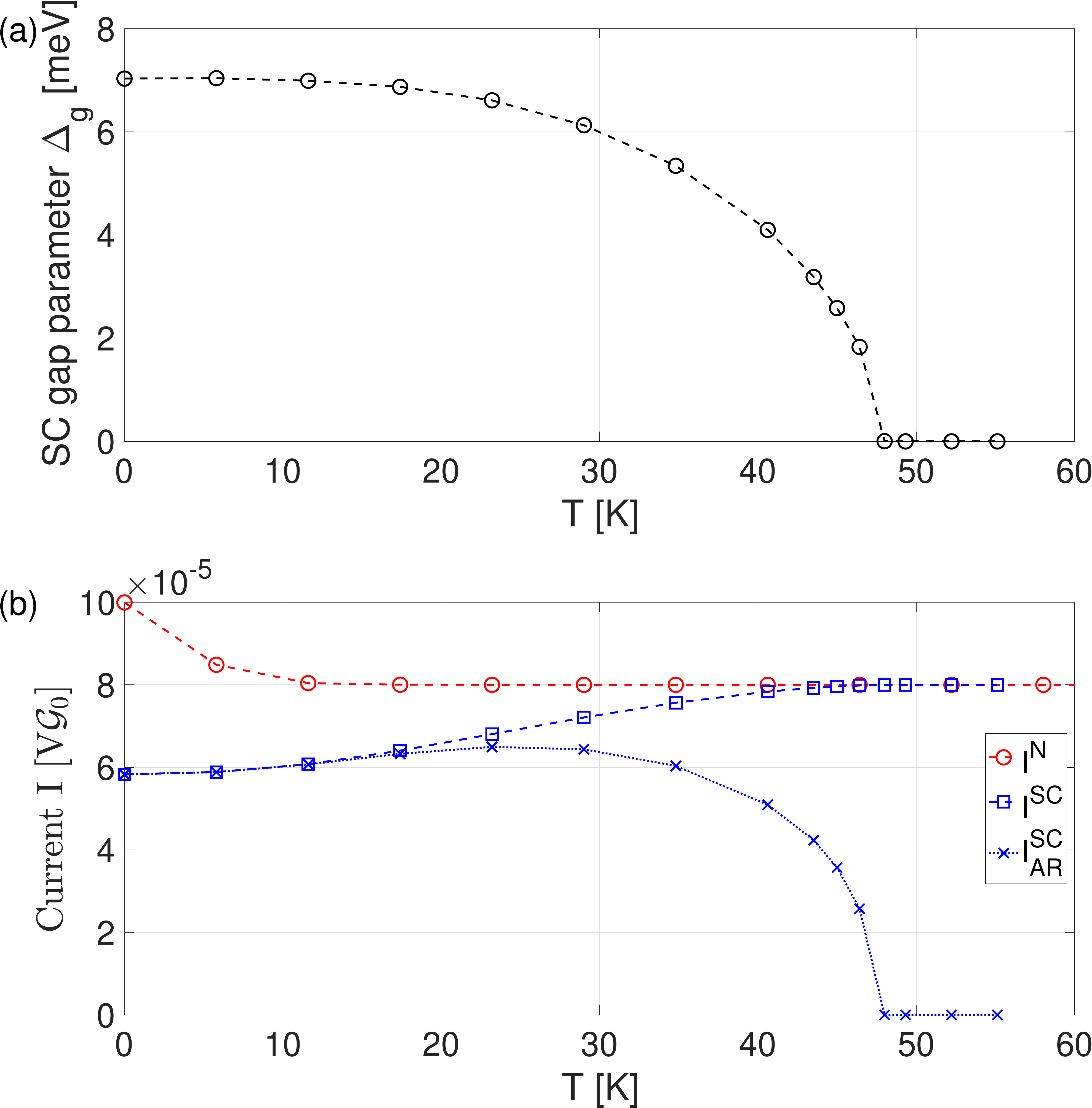}
\caption{Temperature dependence of: a) SC gap parameter $\Delta_g$, b) Current. Device parameters: $N=1001$, $\Gamma=0.25$ eV, $V_{sd}=0.1$ meV, $U=-0.5$ eV. }
\label{fig_T_N1001}
\vspace{-0.0cm}
\end{figure}

The SC gap parameter $\Delta_g \equiv \frac{U}{N} \sum_m |F_{m}|$ is directly related to the magnitude of the gap $E_g$ seen in Fig.~\ref{fig_DOS_N1001} in the electronic DOS of the SC phase, {\it i.e.~}{$\Delta_g\approx E_g/2$}. The dependence of $\Delta_g$ on $T$ shows the expected BCS \cite{BCS} behavior as seen in Fig.~\ref{fig_T_N1001}(a). In the limit of small bias voltage the channel is in the normal phase for temperatures above the critical temperature $T_c = 48$ K. At $T=0$ the SC gap parameter has a value of $\Delta_g(T=0) = 7$ meV which yields a ratio $\Delta_g(T=0)/k_BT_c =1.65$. This is slightly smaller than the BCS ratio of $1.76$ due to $|F_m|$ being smaller near the normal metal contacts than in the middle of the channel (as seen in Fig.~\ref{fig_Fm_N1001}(a)). One can also estimate the SC coherence length according to BCS theory as $\xi_0\equiv \hbar v_F/\pi\Delta_g(T=0) =2\beta a/\pi\Delta_g(T=0)$, implying that the average $\xi_0$ comprises about $45$ sites, which correlates well with the length scale over which the order parameter decreases exponentially towards the contacts as seen in Fig.~\ref{fig_Fm_N1001}(a).

Figure \ref{fig_T_N1001}(b) shows the $T$ dependence of the expectation value of the current through the device $I$, calculated for a fixed low bias voltage $V_{sd}=0.1$ meV. In the normal phase, the zero-temperature, zero-bias conductance equals the quantum unit of conductance ${{\mathcal G}}_0=2e^2/h$, 
with electrons tunneling resonantly through the channel mid-bandwidth level irrespective of the contact transmissivity/coupling parameter $\Gamma$. As $T$ increases, the energy window through which electrons tunnel increases; for $T\gtrsim 10$ K, the effective DOS contributing to the current gets averaged over several energy levels that are broadened by the coupling to the leads. For $\Gamma=\beta/2$, the resulting averaged DOS is $\approx 80\%$ of the peak DOS (Fig.~\ref{fig_DOS_N1001}), explaining the decrease in current seen in Fig.~\ref{fig_T_N1001}(b). Turning to the SC case, one notes that the current/conductance is smaller than in the normal case. This is due to the non-ideality of the metal contacts. Indeed, for a NSN device at zero-temperature, the zero-bias SC conductance approaches ${\mathcal G}_0$ ({\it i.e.~}, the conductance of each NS interface approaches $2\times{\mathcal G}_0$) only if the electron transmission probability between the contacts and the normal-phase channel approaches unity in a finite window around the Fermi level \cite{Cuevas1,Klapwijk,Sun1,WPan}. Within the wide-band limit (WBL) approximation for the leads this happens when $\Gamma=\beta$. For our less ideal choice $\Gamma=\beta/2$, one finds at $T=0$ that the conductance $I^{SC}/V_{sd}$ is about $0.6 \times {\mathcal G}_0$. As seen in  \ref{fig_T_N1001}b, at $T=0$ the SC current takes place via Andreev reflection (AR), as the normal current component is suppressed given the near perfect DOS gap opening in the bulk of the channel (the cross-Andreev component is negligible due to the large channel length). As the temperature increases the DOS gap gets smaller and the DOS at $E=0$ becomes finite  even when projected towards the middle of the chain, as seen in Fig.~\ref{fig_DOS_N1001}(b). Correspondently the normal component $I^{SC}_{N} \approx I^{SC} - I^{SC}_{AR}$ increases until it completely dominates the current for $T\gtrsim T_c$. 

We note from Fig.~\ref{fig_T_N1001}(b) that the continuous decrease of $I^{SC}_{AR}$ as $T$ approaches $T_c$ results in $I^{SC}$ smoothly approaching $I^{N}$ with no apparent discontinuity in the current or in its derivative $\partial I^{SC}/ \partial T$.
Ideally, we would like a device based on a SC channel to switch between the SC and normal state upon photo-excitation, accompanied by a detectable change in the measured current. 
For relatively small perturbations such us those expected to be induced by a single-photon absorption event, an appreciable change in the current could be obtained if the transition is accompanied by a discontinuous change  in the current. Unfortunately, for this device configuration the current undergoes a smooth transition as $T$ changes from below to above $T_C$ as seen in Fig.~\ref{fig_T_N1001}(b), suggesting that the device may not offer a significant advantage over its normal-phase-only counterpart. 

\begin{figure}
\includegraphics[trim=-0 0 -0 0,clip,width=\columnwidth]{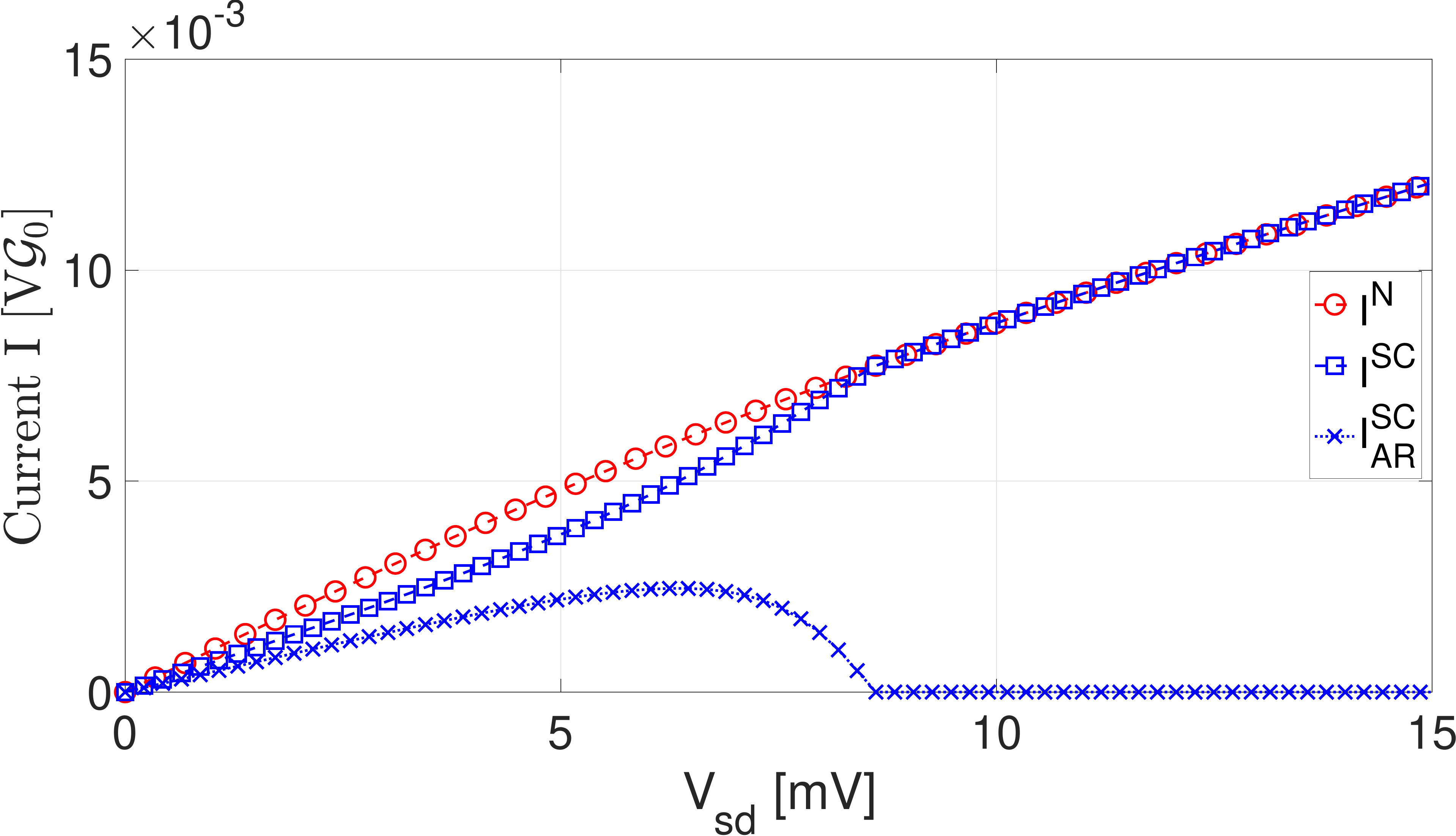}
\caption{Bias voltage dependence of the current in a device driven beyond the critical voltage. Device parameters: $N=201$, $\Gamma=0.25$ eV, $T=0$, $U=-0.5$ eV. }
\label{fig_Vsd_N201}
\vspace{-0.0cm}
\end{figure}

Because in this work we do not consider phonon-based temperature effects that may assist single-photon detection, we focus our attention on the behavior of the device as function of bias voltage.
Figure \ref{fig_Vsd_N201} shows the $I-V$ characteristic of a device  with a smaller channel (we consider $N=201$ to reduce the computational effort), driven beyond the critical voltage/current while the leads are maintained at $T=0$. In the absence of SC the current $I^N$ shows almost linear dependence on the bias voltage $V_{sd}$, {\it i.e.} a behavior close to Ohmic - as expected given the good metal contacts. For low $V_{sd}$ the SC current $I^{SC}$ is about $70\%$ lower than in the normal-phase, and it approaches $I^N$ at the critical voltage $V_c = 8.5$ mV. Translated into temperature, $e/k_B V_c$ yields a value $90\%$ larger than the critical temperature $T_c=55$ K. We note that the Andreev reflection of the current $I^{SC}_{AR}$ drops rather smoothly to zero at $V_{sd}=V_c$; as a consequence the I-V characteristics does not show any discontinuity near the critical voltage. We conclude that for the configurations considered in this section, where the channel is coupled strongly to the electrodes, it is unlikely for the device to undergo a significant change in current due to a small photo-induced perturbation and that a different design strategy is needed.

\subsection{Poor metal contacts/weak channel-electrodes coupling}

A previous study \cite{Snyman} used quasiclassical Green’s functions to show that the I-V characteristics of NSN structures displays, in certain regions of the device parameter space, a discontinuity as the voltage is swept. The behavior of our device as a function of bias voltage suggests this not to be the case when the channel is strongly coupled to the electrodes.  Thus, in this section we consider the opposite case  of weak coupling between channel and electrodes obtainable by setting $\Gamma<<\beta$. In practice, weak channel-electrodes coupling can be realized using few-nm thin insulating materials inserted between the electrode and the channel.

\begin{figure}
\includegraphics[trim=-0 0 -0 0,clip,width=\columnwidth]{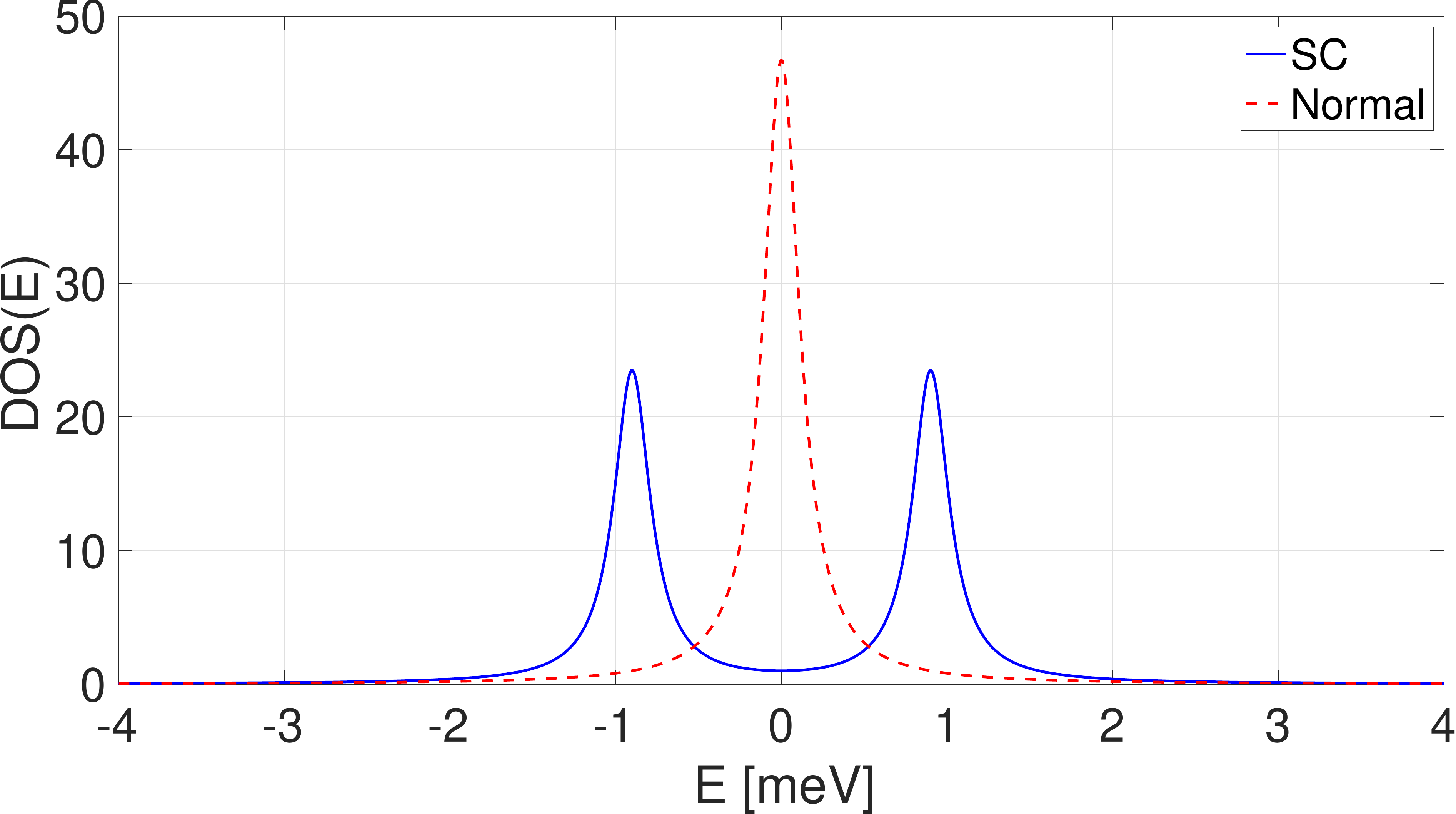}
\caption{Calculated DOS at $T=0$ and $V_{sd}=0$, averaged over all channel sites. Device parameters: $N=51$, $\Gamma=1.7$ meV, $U=-50$ meV.}
\vspace{-0.0cm}
\label{fig_DOS_N51}
\end{figure}

We are interested in a situation \cite{Snyman} where in the absence of SC only one channel level contributes to electronic transport for small bias voltage.  This can be obtained {\it e.g.} by setting $N=51$, $\Gamma=\beta/288\approx 1.7$ meV which yields a separation between channel levels near the Fermi level $\Delta E\approx 60$ meV significantly larger than the level broadening $\sim \Gamma$. Figure \ref{fig_DOS_N51} shows the DOS of such a device in a small energy window near the Fermi level. In the absence of SC, the wavefunction associated with the zero-energy channel level carries its entire weight on the odd sites. We also set $U=-50$ meV such that the resulting SC gap opening in the DOS $E_g = 1.8$ meV is larger than the level broadening but significantly smaller than $\Delta E$.

\begin{figure}
\includegraphics[trim=-0 0 -0 0,clip,width=\columnwidth]{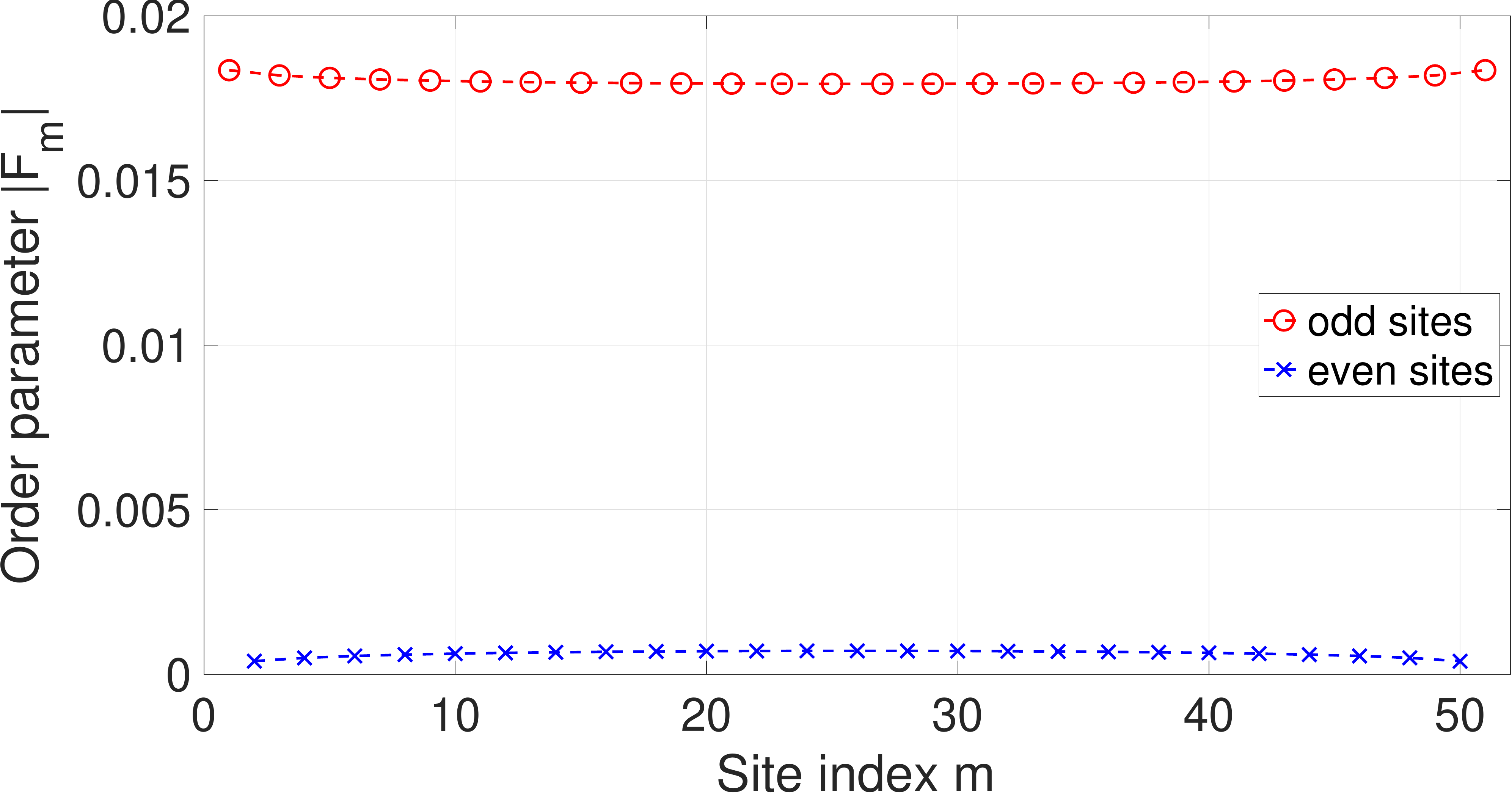}
\caption{Channel SC order parameter $F_m$ at $T=V_{sd}=0$. Device parameters: $N=5$, $\Gamma=1.7$ meV and $U=-50$ meV. }
\label{fig_Fm_N51}
\vspace{-0.0cm}
\end{figure}

Figure \ref{fig_Fm_N51} shows the channel SC order parameter $F_m$ at $T=V_{sd}=0$. We note that the weight of $F_m$ on odd sites is more than an order of magnitude larger than the one on even sites. It is important to note that capturing this feature is not possible without obtaining the SC order parameter self-consistently.
Averaging $F_m$ over all the sites and multiplying by $U$ leads to a SC gap parameter $\Delta_g=0.47$ meV, while averaging only over the odd sites (those that carry the weight of the zero-energy eigen-channel in the absence of SC) one obtains  ${\tilde{\Delta}_g}=0.90$ meV. The latter parameter can be related to the gap opening in the DOS via the usual relationship: $E_g \approx 2\Delta_g^{odd}$.

\begin{figure}
\includegraphics[trim=-0 0 -0 0,clip,width=\columnwidth]{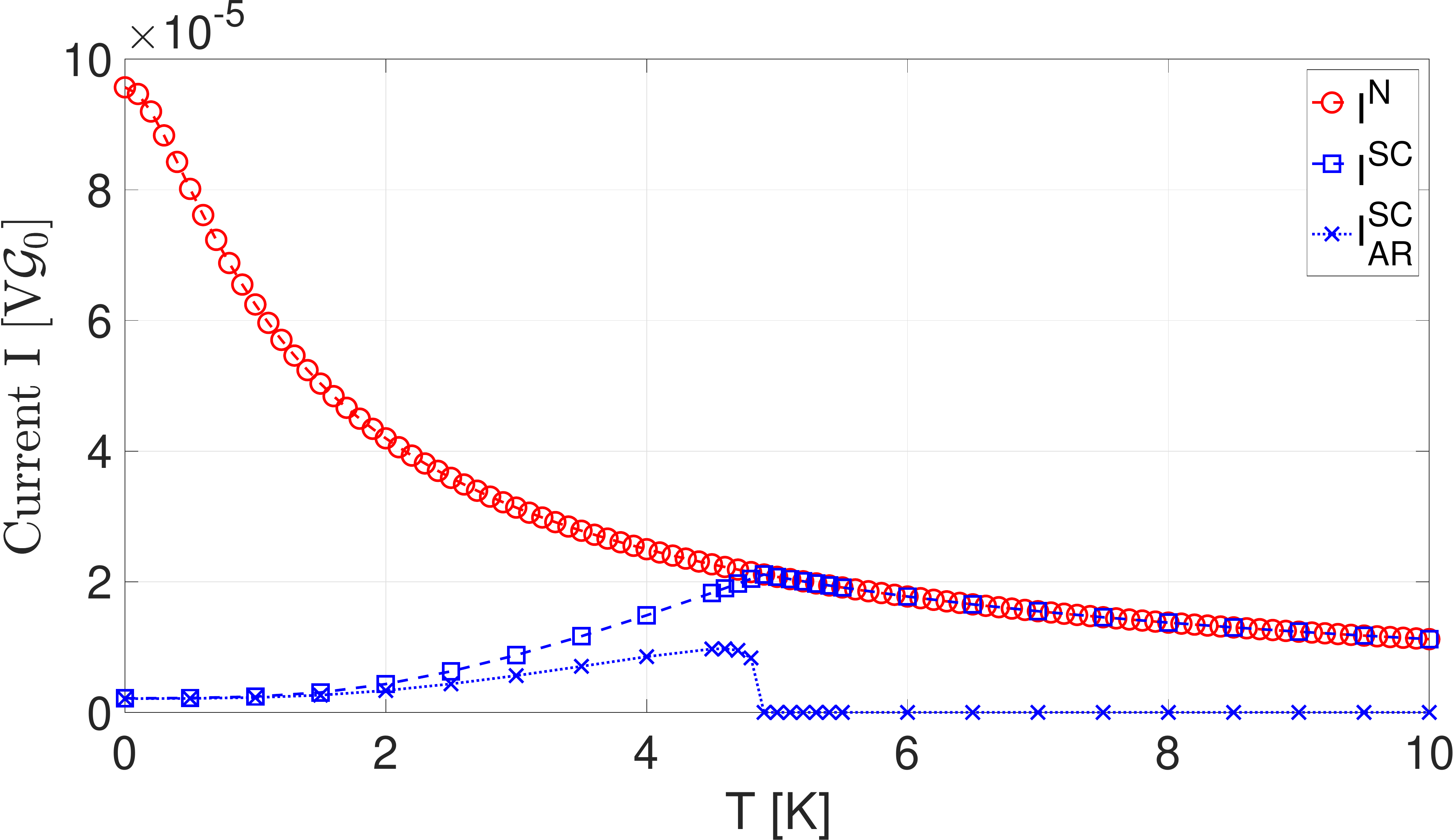}
\caption{Temperature dependence of the current at fixed bias voltage $V_{sd}=0.1$ meV. Device parameters: $N=51$, $\Gamma=1.7$ meV, $U=-50$ meV. }
\label{fig_T_N51}
\vspace{-0.0cm}
\end{figure}

Figure \ref{fig_T_N51} shows the $T$ dependence of the current through the device $I$, calculated for a fixed low bias voltage $V_{sd}=0.1$ meV. As discussed previously, in the normal phase, the zero-temperature, zero-bias conductance equals ${\mathcal G}_0$. A bias voltage $V_{sd}=0.1$ meV is significant enough (w.r.t.~the broadening of the central energy level) to decrease the normal-phase zero-temperature conductance value to $ 0.96 \times {\mathcal G}_0$. Increasing temperature also reduces the normal-phase conductance, this time much more significantly than seen in Fig.~\ref{fig_T_N1001}(b).
Turning to the SC case, we note two differences w.r.t.~to the case depicted in Fig.~\ref{fig_T_N1001}(b). First, at $T=0$, the poor electron transmission between leads and channel results in a conductance {more than $40\times$ smaller than ${\mathcal G}_0$. 
This can be explained by the fact that the effective broadening of the relevant channel electronic level $\tilde{\Gamma} = 0.134$ meV (estimated from the HWHM of the central peak in Fig.~\ref{fig_Fm_N51}) is significantly smaller than the effective gap parameter $\tilde{\Delta}_g=0.9$ meV. Indeed, projecting Eq.~\eqref{I_AR} in the subspace of the only relevant electronic eigen-channel, the zero-bias Andreev reflection conductance reads  (see also Eqs.~\eqref{Gret}, \eqref{SigmaR_nambu} and \eqref{Delta_gap}):
\beq
\lim_{V_{sd}\to 0} \frac{\mathcal{G}_{AR}}{\mathcal{G}_0} \approx \tilde{\Gamma}^2 |G^{r}_{12}(E=0)|^2 
\approx \frac{\tilde{\Gamma}^2 \tilde{\Delta}_g^2}{(\tilde{\Gamma}^2 + \tilde{\Delta}_g^2)^2}, 
\label{G_AR}
\eeq} 
which yields $\mathcal{G}_{AR}\approx\mathcal{G}_0/46$.
Second, as the temperature approaches $T_c = 4.9$ K, while the SC current $I^{SC}$ approaches continuously its normal counterpart $I^N$, its derivative $\partial I^{SC}/ \partial T$ displays a discontinuity at $T=T_c$ due to the sudden drop of the Andreev reflection component of the current $I^{SC}_{AR}$. This discontinuity suggests that the I-V characteristics of the device with poor metal contacts may be fundamentally different than in the good metal contacts case and perhaps it may show more sensitivity to photo-induced perturbations. 

\subsubsection{Impact of small electrostatic gating and magnetic perturbations} 
We consider two mechanisms that can lead to photodetection:
i) a change in the absorber permanent electric dipole moment \cite{Spataru_singlePh,Leonard_SciRep}, {\it e.g.~}as in functionalization with chromophores. This electric dipole change generates an electrostatic potential that acts as a gating potential for the channel. 
For simplicity we mimic this situation with a uniform gating electrostatic potential $V_G$ as the prototype electrostatic perturbation, but we have checked that non-uniform potentials \cite{non-unif_pot} lead to similar conclusions. 
ii) a change in the absorber's permanent magnetic dipole moment $\mu_{abs}$, {\it e.g.~}as in functionalization with small magnetic molecules. We also use a uniform magnetic field $B_0$ to perform an initial assessment of the impact of magnetic perturbations. 

\begin{figure}
\includegraphics[trim=-0 0 -0 0,clip,width=\columnwidth]{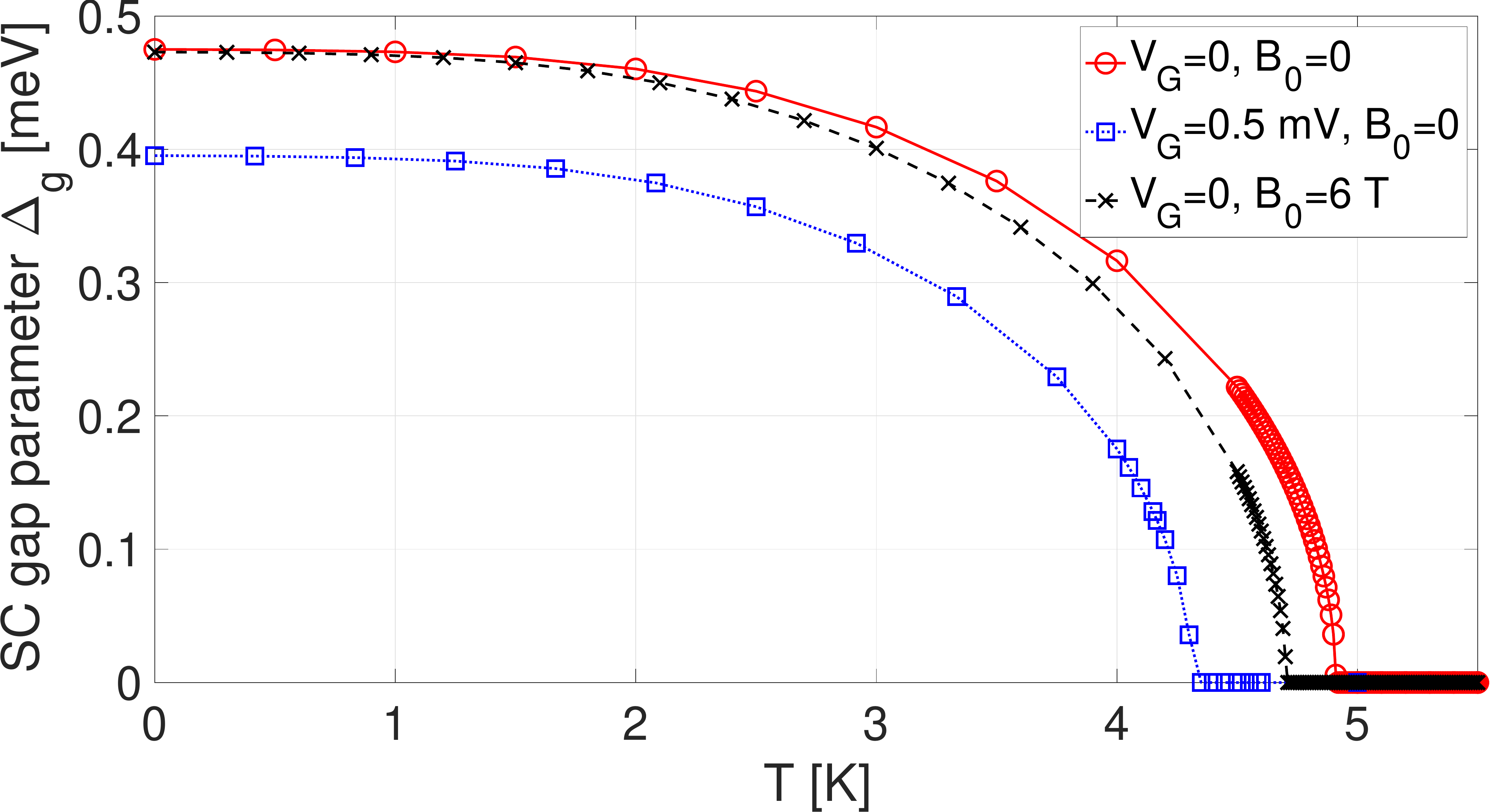}
\caption{Temperature dependence of the SC gap parameter $\Delta_g$ in the absence/presence of several external perturbations such as a uniform gating potential $V_G=0.5$ mV or a uniform magnetic field $B_0=6$ T.  Device parameters: $N=51$, $\Gamma=1.7$ meV, $U=-50$ meV. }
\label{fig_gapPar_T_N51}
\vspace{-0.0cm}
\end{figure}

We start our analysis by studying the behavior of the  SC gap parameter $\Delta_g$ as function of temperature, as shown in Fig.~\ref{fig_gapPar_T_N51}.  In the absence of an electrostatic gating or magnetic perturbation, $\Delta_g(T)$ shows a similar behavior as in the case of good metal contacts (compare to Fig.~\ref{fig_T_N1001}).
Adding a relatively small electrostatic gating potential  $V_G=0.5$ mV yields a reduction in both $\Delta_g(T=0)$ and $T_C$, while a uniform magnetic field $B_0=6$ T reduces $T_C$ by a fraction of a Kelvin but has negligible impact on $\Delta_g(T=0)$. The reduction of $T_c$ may be explained by the fact that in the normal phase both types of perturbation result in a decrease of the DOS at the Fermi level.

\begin{figure}
\includegraphics[trim=-0 0 -0 0,clip,width=\columnwidth]{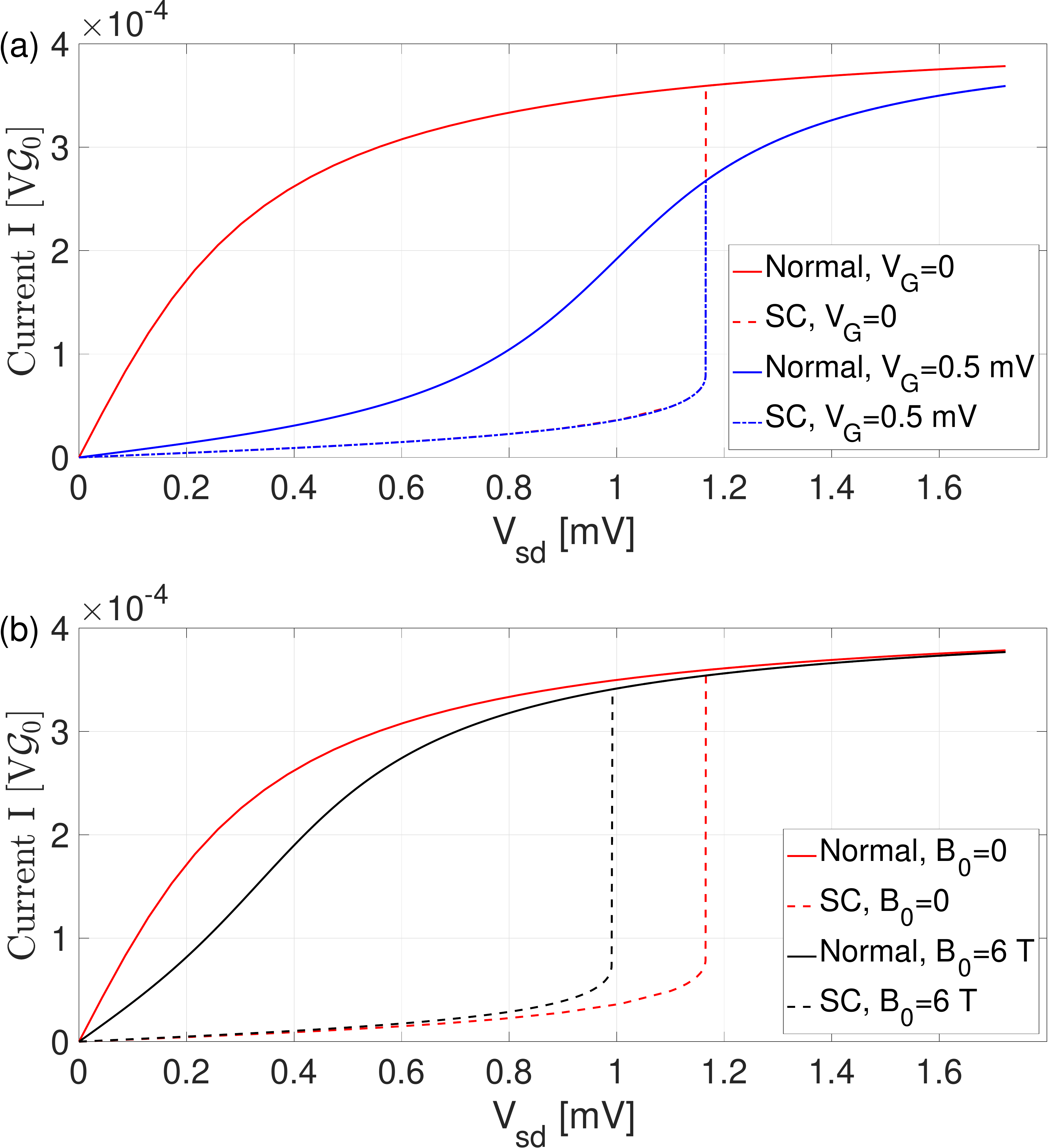}
\caption{Bias voltage dependence of the current at $T=0$ for two different types of perturbations: a) Uniform gating potential $V_G=0.5$ mV b) Uniform magnetic field $B_0=6$ T. Device parameters: $N=51$, $\Gamma=1.7$ meV, $U=-50$ meV.}
\label{fig_Vsd_VG_B0_N51}
\vspace{-0.0cm}
\end{figure}

Based on the above results one could expect that an electrostatic gating potential has a greater impact on the I-V device characteristics than a magnetic field. As discussed next, the results shown in Fig.~\ref{fig_Vsd_VG_B0_N51} indicate that this is not the case. First, in the absence of any external perturbation, as seen in Figs \ref{fig_Vsd_VG_B0_N51}(a) and \ref{fig_Vsd_VG_B0_N51}(b), the normal-phase current $I^N$ shows non-linear behavior due to the non-Ohmic character of the metal contacts. In fact $I^N$ develops a plateau when $V_{sd}$ increases beyond the resonant-level broadening. Turning to the SC case, an important feature is that the current $I^{SC}$ undergoes a sharp jump near the critical voltage $V_c= 1.17$ mV. The jump $I^N-I^{SC}$ is significant, amounting to $(I^N-I^{SC})/I^{SC} = 4.5$.

Figure \ref{fig_Vsd_VG_B0_N51}(a) shows the impact of the electrostatic gating potential $V_G=0.5$ mV on the  I-V characteristics. Because the channel level through which electrons tunnel is off-resonance, the normal-phase current is much reduced w.r.t.~the  $V_G=0$ case at $V_{sd} \gtrsim 0$. As $V_{sd}$ increases beyond both the level broadening and  $V_G$, $I^N$ starts approaching the unperturbed values. We find that as long as the channel is in the SC phase, for $V_{sd}<V_c$, the electrostatic gating potential $V_G$ has no impact on the current $I^{SC}$ or the critical voltage $V_c$. This is a somewhat unexpected finding given that $V_G=0.5$ mV does have an appreciable impact on $T_c$ as seen in Fig.~\ref{fig_gapPar_T_N51}. While all the results in Fig.~\ref{fig_Vsd_VG_B0_N51} are obtained with metal contacts maintained at $T=0$, we have tried other temperatures but always reached the same result, namely that the electrostatic gating mechanism is not effective at improving the light-detection efficiency of a functionalized SC channel w.r.t.~it normal-phase counterpart.

The impact of a uniform magnetic field \cite{surf_currents} $B_0=6$ T is seen in Fig.~\ref{fig_Vsd_VG_B0_N51}(b). In the normal phase, its effect is to reduce the current $I^N$ by a factor of $2.7$ at $V_{sd} \gtrsim 0$ and by about $3\%$ at $V_{sd} \approx 1$ mV. In the SC case, the reduction of $I^{SC}$ is $<15\%$ for $V_{sd}<0.6$ mV. Interestingly, the impact of $B_0$ increases at higher $V_{sd}$ and most importantly it results in an $18\%$ decrease of the critical voltage $V_c$ from $1.17$ mV to $0.99$ mV. 
This suggests that functionalizing a SC with a magnetic molecule might result in an efficient detector if one operates the device near $V_c$ and if the photo-excitation of the molecule induces a local magnetic dipole whose associated magnetic field is strong enough to yield a measurable change in $V_c$.

\begin{figure}
\includegraphics[trim=-0 0 -0 0,clip,width=\columnwidth]{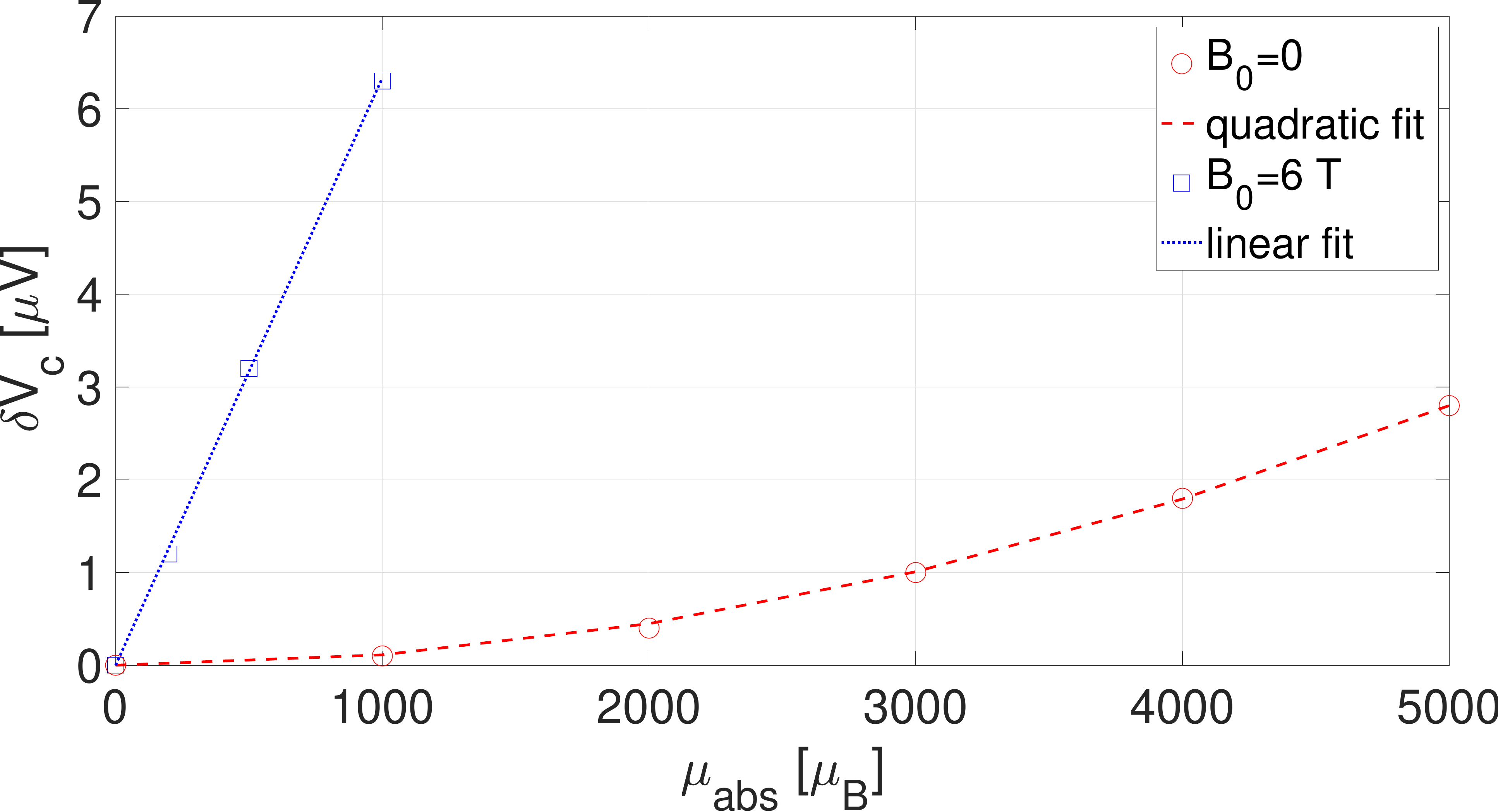}
\caption{The impact of a magnetic dipole moment $\mu_{abs}$ on the critical voltage $\delta V_c$ at $T=0$ in the presence/absence of a background uniform magnetic field $B_0=6$ T.
Device parameters: $N=51$, $\Gamma=1.7$ meV, $U=-50$ meV.}
\label{fig_dVc_mu_N51}
\vspace{-0.0cm}
\end{figure}

To estimate the impact of a photo-excited magnetic absorber on the critical voltage $V_c$ we consider a magnetic dipole characterized by a magnetic moment $\mu_{abs}$ and positioned at a distance $d=1$ nm away from the central atom of the SC channel. The magnetic dipole is oriented perpendicular to the plane defined by the channel and the line joining the magnetic dipole and the mid-channel site. The magnetic potential field at the channel site 'm' situated at a distance $r_m=\sqrt{a^2(m-m_0)^2+d^2}$ from the mid-channel site $m_0=(N+1)/2$ is $ B^{abs}_m=\frac{\mu_0}{4\pi }\frac{\mu_{abs}}{r_m^3}$ with $\mu_0$ the vacuum permeability, and the corresponding Zeeman splitting potential $V_Z$ acting on spin up/down electrons in the chain is \cite{spin_suscep} $V_Z(m)=\pm \mu_B B^{abs}_m/2$.

We calculate the I-V characteristics in the vicinity of $V_c$ for various $\mu_{abs}$ and extract the change $\delta V_c \equiv V_c(\mu_{abs})-V_c(\mu_{abs}=0)$, as plotted in Fig.~\ref{fig_dVc_mu_N51} with red circles. As expected from the symmetry of the problem w.r.t. a change in sign of the magnetic field,  we find that $\delta V_c$ depends quadratically on $\mu_{abs}$: $\delta V_c \sim \mu_{abs}^2$, as indicated by the quadratic fit plotted with the red, dashed line. Quantitatively, the change $\delta V_c$ is on the order of $0.1\mu$V for $\mu_{abs}$ of about $1000\times \mu_B$. This large change in $\mu_{abs}$ may be difficult to achieve for single photon absorption.

However, the quadratic behavior also implies that the presence of a background magnetic field $B_0$ significantly larger than the one produced by $\mu_{abs}$ can amplify the impact of an added magnetic moment: $V_c(\mu_{abs};B_0)-V_c(\mu_{abs}=0;B_0)\sim  \mu_{abs} B_0 $. Indeed, Fig.~\ref{fig_dVc_mu_N51} shows that in the presence of a background uniform magnetic field $B_0=6$ T, the change $\delta V_c$ shows a linear dependence on $\mu_{abs}$ with a slope of $0.0063$ $\mu V/\mu_B$. Translated into temperature, one obtains that a change $e/k_B \delta V_c=10$ mK can be obtained by an excited absorber magnetic moment $\mu_{abs}=150 \times \mu_B$. 

\begin{figure}
\includegraphics[trim=-0 0 -0 0,clip,width=\columnwidth]{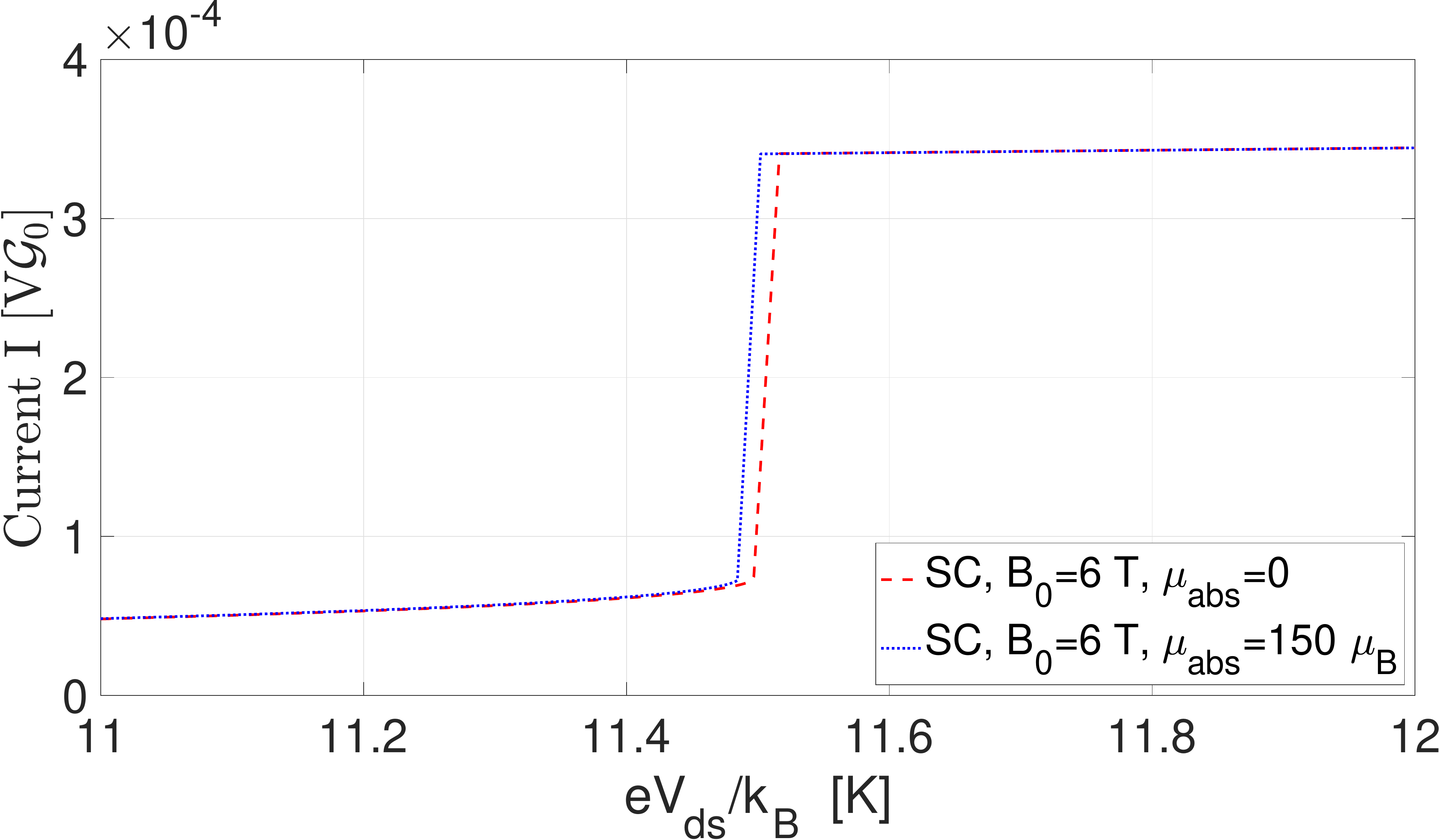}
\caption{The I-V characteristics of the device at $T=0$ in the presence of a background uniform magnetic field $B_0=6$ T and without/with a magnetic dipole moment $\mu_{abs}=150 \times \mu_B$.
Device parameters: $N=51$, $\Gamma=1.7$ meV, $U=-50$ meV.}
\label{fig_IV_mu150_N51}
\vspace{-0.0cm}
\end{figure}

Figure \ref{fig_IV_mu150_N51} shows the I-V characteristics of the device in the presence of a background uniform magnetic field  $B_0=6$ T. We consider a magnetic molecule that has no magnetic dipole in the ground state but carries a magnetic moment $\mu_{abs}=150 \times \mu_B$ in the excited state. As noted previously, the change in the critical voltage is about $10$ mK upon photo-excitation of the magnetic molecule. Controlling the voltage within $10$ mK is experimentally feasible, which means that if one operates the device while maintaining the bias voltage within $10$ mK below the critical voltage, photoexcitation of the magnetic molecule would trigger the SC-to-normal phase transition of the channel. As seen in Fig. \ref{fig_IV_mu150_N51}, this transition is accompanied by an increase of the current by a factor $>5$, {\it i.e.} 
orders of magnitude higher than one would obtain with a normal-phase channel and otherwise similar device parameters. For comparison, we found that the current through a {\it semiconducting} but otherwise similar electronic transport channel may increase, due to a photo-induced $10$ Debye electrical dipole situated about $1$ nm away from the channel, by a factor of $1$ at most upon device optimization \cite{Spataru_singlePh}.

\subsection{Conclusion}

We have studied a photon detector device that consists of a functionalized SC transport channel contacted by two normal-phase metal electrodes. We have considered two possible detection mechanisms based on i) electrostatic gating effects ({\it e.g.} functionalization with chromophores) and ii) magnetic effects ({\it e.g.} functionalization with magnetic molecules). We found that optimal device designs require weak coupling between the leads and the channel $\Gamma $ ({\it e.g.} using insulating layers to separate the metal contacts from the channel) such that only one channel electronic level participates in transport and the I-V device characteristics shows a discontinuity near the critical voltage. Our results indicate that the electrostatic gating mechanism is not effective for triggering a SC-to-normal transition. However, we find that magnetic effects offer an efficient photo-detection mechanism whereas in the presence of a background magnetic field, photo-excitation of a magnetic molecule may trigger a SC-to-normal transition accompanied by a measurable change in the device current.

\renewcommand{\theequation}{A-\arabic{equation}} 
\setcounter{equation}{0}
\setcounter{subsection}{0}

\section*{Appendix: Self-consistent Keldysh-Nambu NEGF formalism}
To account for superconductivity and non-equilibrium phenomena the relevant Keldysh Green's function is defined using $4$-component Nambu spinors:
\begin{multline}
\check{G}_{m,n}(t,t') \\
=
-i\langle T_C \left\{
\left(
\begin{aligned}[c]
c_{\uparrow m}(t)\\
c^\dagger_{\downarrow m}(t)\\
c_{\downarrow m}(t)\\
c^\dagger_{\uparrow m}(t)
\end{aligned}
\right)
\otimes
\left(
\begin{aligned}[l]
c^\dagger_{\uparrow n}(t') , c_{\downarrow n}(t'), c^\dagger_{\downarrow n}(t') , c_{\uparrow n}(t')
\end{aligned}
\right)
\right\}\rangle 
\end{multline}
where $T_C$  denotes time ordering along the double-time Keldysh contour.

For BCS-type superconductivity with coupling between opposite spins and in the presence of either spin or electron-hole symmetry (which are the situations considered in this work), we may use only 2-component Nambu spinors. Neglecting for the sake of simplicity spin indices (the spin structure is $\uparrow\uparrow$,$\uparrow\downarrow$,$\uparrow\downarrow$,$\downarrow\downarrow$ for the $_{11}$,$_{12}$,$_{21}$,$_{22}$ Nambu components respectively), the Keldysh Green's function is written in the reduced $2 \times 2$  Nambu space: 
\begin{multline}
\check{G}_{m,n}(t,t')=-i\langle T_C \left\{
\left(
\begin{aligned}[c]
c_m(t)\\
c^\dagger_m(t)
\end{aligned}
\right)
\otimes
\left(
\begin{aligned}[l]
c^\dagger_n(t') , c_n(t')
\end{aligned}
\right)
\right\}\rangle \\
=-i\langle T_C \left\{
\left(
\begin{aligned}[c]
c_m(t)c^\dagger_n(t') , c_m(t)c_n(t')
\\
c^\dagger_m(t)c^\dagger_n(t') , c^\dagger_m(t)c_n(t')
\end{aligned}
\right)
\right\}\rangle.
\end{multline} 

The corresponding Keldysh components lesser ($<$), greater ($>$), advanced ($a$), and retarded ($r$) of the Green's function are then:
\begin{multline}
\hat{G}^<_{m,n}(t,t')=+i 
\left(
\begin{aligned}[c]
\langle c^\dagger_n(t')c_m(t)\rangle , \langle c_n(t')c_m(t)\rangle
\\
\langle c^\dagger_n(t')c^\dagger_m(t)\rangle , \langle c_n(t')c^\dagger_m(t)\rangle
\end{aligned}
\right) \\
\equiv
\left(
\begin{aligned}[c]
G^<_{m,n}(t,t') , \ F_{m,n}(t,t')
\\
\bar{F}_{m,n}(t,t') , -G^>_{n,m}(t',t)
\end{aligned}
\right),
\end{multline} 

\begin{multline}
\hat{G}^>_{m,n}(t,t')=-i 
\left(
\begin{aligned}[c]
\langle c_m(t) c^\dagger_n(t')\rangle , \langle c_m(t) c_n(t')\rangle
\\
\langle c^\dagger_m(t) c^\dagger_n(t')\rangle , \langle c^\dagger_m(t) c_n(t')\rangle
\end{aligned}
\right) \\
\equiv
\left(
\begin{aligned}[c]
G^>_{m,n}(t,t') , \ \bar{F}^*_{m,n}(t,t')
\\
F^*_{m,n}(t,t') , -G^<_{n,m}(t',t)
\end{aligned}
\right),
\end{multline}

\begin{multline}
\hat{G}^r_{m,n}(t,t')=\theta(t-t')[\hat{G}^>_{m,n}(t,t')-\hat{G}^<_{m,n}(t,t')] \\
\equiv 
\left(
\begin{aligned}[c]
G^{r}_{m,n}(t,t') , \ \theta(t-t')[\bar{F}^*_{m,n}(t,t')-F_{m,n}(t,t')]
\\
\theta(t-t')[F^*_{m,n}(t,t')-\bar{F}_{m,n}(t,t')] , -G^{a}_{n,m}(t',t)
\end{aligned}
\right),
\end{multline}

\begin{multline}
\hat{G}^a_{m,n}(t,t')=\theta(t'-t)[\hat{G}^<_{m,n}(t,t')-\hat{G}^>_{m,n}(t,t')] \\
\equiv 
\left(
\begin{aligned}[c]
G^{a}_{m,n}(t,t') , \ -\theta(t'-t)[\bar{F}^*_{m,n}(t,t')-F_{m,n}(t,t')]
\\
-\theta(t'-t)[F^*_{m,n}(t,t')-\bar{F}_{m,n}(t,t')] , -G^{r}_{n,m}(t',t)
\end{aligned}
\right).
\end{multline} 

If we reference all single-particle energies as well as the lead chemical potentials w.r.t. the condensate chemical potential $\mu_C$ when superconductivity is
present, then at steady-state one can replace the double-time dependencies with the time difference $(t,t')\rightarrow (t-t')$. After Fourier-transform we obtain:
\begin{multline}
\hat{G}^<_{m,n}(\omega)\equiv
\left(
\begin{aligned}[c]
G^<_{m,n}(\omega) \ , \ F_{m,n}(\omega) \ \ 
\\
\bar{F}_{m,n}(\omega) \ , -G^>_{n,m}(-\omega)
\end{aligned}
\right),
\end{multline} 
\begin{multline}
\hat{G}^r_{m,n}(\omega)\equiv
\left(
\begin{aligned}[c]
G^r_{m,n}(\omega) \ , \ G^{r,12}_{m,n}(\omega) \ \ 
\\
G^{r,21}_{m,n}(\omega) \ , -G^a_{n,m}(-\omega)
\end{aligned}
\right),
\end{multline} 
and similar for the other Keldysh components. Note that $\bar{F}_{m,n}(t,t')=-{F}^*_{n,m}(t',t)$ [or $\bar{F}(\omega)=-{F}(\omega)^\dagger$] so one can write (T stands for transpose):
\begin{multline}
\hat{G}^<(\omega)=
\left(
\begin{aligned}[c]
G^<(\omega) \,\,\,\, \ ,  \ F(\omega) \ \ 
\\
-F^\dagger(\omega) \ , -G^{>,T}(-\omega)
\end{aligned}
\right) , \hat{G}^<(\omega)=-{\hat{G}^<(\omega)}^\dagger
\end{multline}

and

\begin{multline}
\hat{G}^>(\omega)=
\left(
\begin{aligned}[c]
G^>(\omega) \,\,\,\, \ ,  \ F^\dagger(\omega) \ \ 
\\
-F(\omega) \ , -G^{<,T}(-\omega)
\end{aligned}
\right) , \hat{G}^>(\omega)=-{\hat{G}^>(\omega)}^\dagger.
\end{multline}

To calculate the Green's functions we approximate self-energies at the Hartree-Fock level. 

The non-interacting, isolated channel Green's function is:
\begin{multline}
\hat{g}^r_{m,n}(\omega)
=\left(
\begin{aligned}[c]
{g}^r_{m,n}(\omega) \  , \ \ \ 0 \ \
\\
0 \ , -{g}^a_{n,m}(-\omega)
\end{aligned}
\right)
=\left(
\begin{aligned}[c]
{g}^r_{m,n}(\omega) \  , \ \ \ 0 \ \
\\
0 \ , -{g}^a_{m,n}(-\omega)
\end{aligned}
\right).
\end{multline} 
Here we used that $g^{r,a}$ is symmetric because the channel Hamiltonian $H^C$ is symmetric with real matrix elements and real eigenvectors $v_j$. Then $g^a_{n,m}=v(n,j)*g^a_{j,j}*v'(j,m)=v(n,j)*g^a_{j,j}*v(m,j)=g^a_{m,n}$.

In the channel eigenstate basis (indexed by $j$), one has:
\beq
{g}^r_{j,j}(\omega)=\frac{1}{\omega-(\epsilon_j-V_G)+i\delta}
\eeq
and
\begin{multline}
-{g}^a_{j,j}(-\omega)=-\frac{1}{-\omega-(\epsilon_j-V_G)-i\delta} \\
=\frac{1}{\omega+(\epsilon_j-V_G)+i\delta}.
\end{multline}

The non-interacting, isolated leads Green's functions, are written in the leads eigenstate basis (indexed by $k_{L,R}$) as:
\begin{multline}
\hat{g}^{L,R,r}_{k_{L,R},k_{L,R}}(\omega)=
\left(
\begin{aligned}[c]
{g}^{L,R,r}_{k_{L,R},k_{L,R}}(\omega) \  , \ \ \ 0 \ \
\\
0 \ , -{g}^{L,R,a}_{k_{L,R},k_{L,R}}(-\omega)
\end{aligned}
\right).
\end{multline} 
where:
\beq
{g}^{L,R,r}_{k_{L,R},k_{L,R}}(\omega)=\frac{1}{\omega-\epsilon^{L,R}_k+i\delta}
\eeq
and
\begin{multline}
-{g}^{L,R,a}_{k_{L,R},k_{L,R}}(-\omega)=-\frac{1}{-\omega-\epsilon^{L,R}_k-i\delta} \\
=\frac{1}{\omega+\epsilon^{L,R}_k+i\delta}.
\end{multline}

Also,
\begin{multline}
\hat{g}^{L,R,<}_{k_{L,R},k_{L,R}}(\omega)=
\left(
\begin{aligned}[c]
{g}^{L,R,<}_{k_{L,R},k_{L,R}}(\omega) \  , \ \ \ 0 \ \
\\
0 \ , -{g}^{L,R,>}_{k_{L,R},k_{L,R}}(-\omega)
\end{aligned}
\right)
\end{multline} 
with:
\begin{multline}
{g}^{L,R,<}_{k_{L,R},k_{L,R}}(\omega)=2\pi i f(\epsilon^{L,R}_k-\mu_{L,R}) \delta(\omega-\epsilon^{L,R}_k)= \\
2\pi i f(\omega-\mu_{L,R}) \delta(\omega-\epsilon^{L,R}_k)
\end{multline}
and
\begin{multline}
-{g}^{L,R,>}_{k_{L,R},k_{L,R}}(-\omega)=2\pi i [1-f(\epsilon^{L,R}_k-\mu_{L,R})] \delta(-\omega-\epsilon^{L,R}_k) \\
= 2\pi i [1-f(-\omega-\mu_{L,R})] \delta(\omega+\epsilon^{L,R}_k) \\
= 2\pi i f(\omega+\mu_{L,R}) \delta(\omega+\epsilon^{L,R}_k).
\end{multline}
where $f$ is the Fermi-Dirac distribution function $f(E)=1/[exp(E/k_BT) + 1]$ and we used that $1-f(-E)=f(E)$ . We emphasize that all single-particle energies as well as the chemical potentials of the leads must be referenced w.r.t. the chemical potential of the SC condensate $\mu_C$ in order for the above expressions to be valid when superconductivity is present. 
$\mu_C$  can be determined self-consistently by imposing current conservation \cite{Lambert} along the channel. For the case where the total system has electron-hole symmetry one has $\mu_C=0$.

The leads are treated within WBL in which case the retarded lead self-energies are purely imaginary:
\begin{multline}
\hat{\Gamma}^{L,R,r}_{m,n}(\omega)= \\
\sum_{k_{L,R}}T_{m,k_{L,R}} \frac{\hat{g}^{L,R,a}_{k_{L,R},k_{L,R}}(\omega)-\hat{g}^{L,R,r}_{k_{L,R},k_{L,R}}(\omega)}{2i} T_{k_{L,R},n}
= \\
\left(
\begin{aligned}[c]
\Gamma^{L,R}_{m,n}(\omega),  0 
\\
0, \Gamma^{L,R}_{m,n}(-\omega)
\end{aligned}
\right)
\eqWBL
\left(
\begin{aligned}[c]
\Gamma^{L,R}_{m,n},  0 
\\
0, \Gamma^{L,R}_{m,n}
\end{aligned}
\right)
\end{multline}
where ($T$ is the leads-channel tunneling Hamiltonian):
\begin{multline}
\Gamma^{L,R}_{m,n}(\omega)\equiv \pi \sum_{k_{L,R}}T_{m,k_{L,R}} T_{k_{L,R},n} \delta(\omega-\epsilon^{L,R}_k)
\end{multline}
while the lesser lead self-energies are:
\begin{multline}
\hat{\Gamma}^{L,R,<}_{m,n}(\omega)=\sum_{k_{L,R}}T_{m,k_{L,R}} \hat{g}^{L,R,<}_{k_{L,R},k_{L,R}}(\omega)T_{k_{L,R},n} \\
\eqWBL
2i \left(
\begin{aligned}[c]
f(\omega-\mu_{L,R})\Gamma^{L,R}_{m,n},  0 
\\
0, f(\omega+\mu_{L,R})\Gamma^{L,R}_{m,n}
\end{aligned}
\right).
\label{Gammaless}
\end{multline} 

The channel Green's functions are obtained via the equations:

\beq
\hat{G}^<(\omega)=\hat{G}^r(\omega)
\left[\hat{\Gamma}^{L,<}+\hat{\Gamma}^{R,<}\right]\hat{G}^a(\omega),
\label{Gless}
\eeq

\beq
\hat{G}^r(\omega)=\frac{1}{
{\hat{g}^r(\omega)}^{-1} -\hat{\Sigma}^r
+i\hat{\Gamma}^L+i\hat{\Gamma}^R,
}
\label{Gret}
\eeq

and

\beq
\hat{G}^a(\omega)={\hat{G}^r(\omega)}^\dagger.
\eeq

The SC pairing potential inside the channel has the form: $\hat{U}^{SC}_{m,n}=U^{SC}_{m,n}\hat{\tau}_1$ ($\hat{\tau}_i$ are the Pauli matrices). This interaction has zero contribution to the Hartree self-energy diagram while the channel Fock self-energy $\hat{\Sigma}^r$ is obtained self-consistently via:
\begin{multline}
\hat{\Sigma}^r_{m,n} = 
i \hat{U}^{SC}_{m,n} .* \\
\int \frac{d \omega}{2\pi} 
\hat{\tau}_3 * \left[\hat{G}^<_{m,n}(\omega) + \hat{G}^r_{m,n}(\omega)*e^{i\hat{\tau}_3 \delta \omega}
\right] * \hat{\tau}_3 \\
=
-  U^{SC}_{m,n} 
\left(
\begin{aligned}[c]
0 , \int \frac{d \omega}{2\pi i} [F_{m,n}(\omega) -G^{r,12}_{m,n}(\omega)]\\
\int \frac{d \omega}{2\pi i} [\bar{F}_{m,n}(\omega)-G^{r,21}_{m,n}(\omega)] , 0 
\end{aligned}
\right).
\end{multline}
where $.*$ stands for element-by-element multiplication in Nambu space.
We note that the integral $\int \frac{d \omega}{2\pi i} G^{r,12}_{m,n}(\omega)$ vanishes as it is equal to $G^{r,12}_{m,n}(t,t)=\theta(0)[\bar{F}^*_{m,n}(t,t)-F_{m,n}(t,t)]=i/2\langle \{c_n(t),c_m(t)\}\rangle$ which is zero by virtue of equal-time fermion anti-commutation rules.
 
For the SC pairing potential one chooses an attractive, contact interaction $U^{SC}_{m,n}=-\delta_{m,n} U$ (see the Hubbard-like term of $H^C$ in eq. \eqref{H_ch}) which leads to:
\begin{multline}
\hat{\Sigma}^r_{m}=-U
\left(
\begin{aligned}[c]
0 , \int \frac{d \omega}{2\pi i} F_{m,m}(\omega) \\
\int \frac{d \omega}{2\pi i} \bar{F}_{m,m}(\omega) , 0 
\end{aligned}
\right) \\
=
-U 
\left(
\begin{aligned}[c]
0 , <c_m(t)c_m(t)> \\
<c_m^\dagger(t)c_m^\dagger(t)> , 0 
\end{aligned}
\right)
\equiv
-U 
\left(
\begin{aligned}[c]
0 , F_{m} \\
F^*_{m} , 0 
\end{aligned}
\right)
\label{SigmaR_nambu}
\end{multline}

The SC gap parameter is defined as $U$ times the average of the SC order parameter over the  channel sites: 
\beq
\Delta_g \equiv \frac{U}{N} \sum_m |F_{m}|.
\label{Delta_gap}
\eeq

\subsection{Current expression}
To obtain the expression for the current one starts from the usual expression \cite{Wingreen,Spataru_AM1,Claro} for the current through the left contact
\begin{multline}
I_L
=i\frac{e}{h} \int d \omega Tr\{\left[\hat{\Gamma}^{L,r} \hat{G}^<(\omega)+\hat{\Gamma}^{L,<} \frac{\hat{G}^r(\omega)-\hat{G}^a(\omega)}{2i}\right]_{11}\} \\
=i\frac{e}{h} \int d \omega Tr\{\Gamma^L {G}^<(\omega)-2if(\omega-\mu_L)\Gamma^L \\
\times \left[\hat{G}^r(\omega)(\hat{\Gamma}^L+\hat{\Gamma}^R)\hat{G}^a(\omega)\right]_{11}\}
\end{multline}
where $[..]_{11}$ stands for the upper diagonal Nambu component. In the last equality we made use of Eqs.~\eqref{Gammaless}-\eqref{Gret}. 

Further use of Eq.~\eqref{Gless} as well as permutation properties of symmetric matrices (inside the trace) yields: 
\beq
I_L=I_{NT}+I_{AR}+I_{CA}
\eeq
where the normal transmission (NT) component is:
\begin{multline}
I_{NT} = 2 \frac{e}{h} \int d \omega [f(\omega-\mu_L)-f(\omega-\mu_R)] \\
\times Tr\{\Gamma^L G^{r,11}(\omega) \Gamma^R G^{a,11}(\omega)\}
\end{multline}
and the Andreev reflection (AR) / cross-Andreev (CA) components read:
\begin{multline}
I_{AR} = 2 \frac{e}{h} \int d \omega [f(\omega-\mu_L)-f(\omega+\mu_L)] \\
\times Tr\{\Gamma^L G^{r,12}(\omega) \Gamma^L G^{a,21}(\omega)\},
\end{multline}
\begin{multline}
I_{CA} = 2 \frac{e}{h} \int d \omega [f(\omega-\mu_L)-f(\omega+\mu_R)] \\
\times Tr\{\Gamma^L G^{r,12}(\omega) \Gamma^R G^{a,21}(\omega)\},
\end{multline}
with $G^{r/a,ij} \equiv \left[ \hat{G}^{r/a}\right]_{ij}$ being the $ij$ Nambu component of the retarded/advanced Green's functions. We note that the cross-Andreev component of the current vanishes in the case where the lead chemical potentials are symmetric w.r.t. SC condensate chemical potential ({\it i.e.} $\mu_L=-\mu_R=V_{sd}/2$).

A similar expression holds for the current through the right contact $I_R$ (for the steady-state situations considered in this paper one has $I_L=I_R$).

The expressions for the current can be evaluated numerically as simple matrix multiplications, taking advantage of the fact that the integrals over $\omega$ can be reduced to the form [$z_i$ being the complex eigenvalues of $\omega-\hat{G}^{r^{-1}}(\omega)$]:
\beq
\int d \omega \frac{f(\omega-\mu)}{(\omega-z_i)(\omega-z_j^*)}
\eeq
which is evaluated analytically by performing the exact summation -using the digamma function- over the residues of the Fermi-Dirac distribution function $f$. 
 
\begin{acknowledgments}
Work supported by the Defense Advanced Research Projects Agency (DARPA) DETECT program. The views, opinions and/or findings expressed are those of the author and should not be interpreted as representing the official views or policies of the Department of Defense or the U.S. Government. Sandia National Laboratories is a multi-mission laboratory managed and operated by National Technology and Engineering Solutions of Sandia, LLC., a wholly owned subsidiary of Honeywell International, Inc., for the U.S. Department of Energy's National Nuclear Security Administration under contract DE-NA-0003525. 
\end{acknowledgments}



\begin{thebibliography}{99}
\bibitem{non-unif_pot} We considered diatomic absorbers with permanent dipoles of several Debye positioned about $1$ nm away from the channel, and described the absorber-channel Coulomb interaction within the Ohno parametrization \cite{Ohno,Capaz_trends_Ohno}. We found that such electrical dipoles have negligible impact on the $I-V$ characteristic when  the channel is in the SC phase.
\bibitem{surf_currents} The applied magnetic field is not expulsed from the SC channel because the 'surface' of the atomic chain has zero surface area and thus no internal surface currents could develop to counter the external magnetic field via the Meissner effect. For the same reason we do not expect our results on the impact of magnetic perturbations to hold for nanowires/nanotubes with a diameter larger than the London penetration depth.
\bibitem{spin_suscep} While the channel Hamiltonian \eqref{H_ch} does not account for magnetic susceptibility, we have studied separately the impact of the magnetic field generated by the magnetic moments induced on the channel sites via Zeeman splitting and found it to be negligible.
\end{thebibliography}
\end{document}